\documentclass[conference, final]{IEEEtran}

\usepackage[T1]{fontenc}
\usepackage[english]{babel}
\usepackage[utf8]{inputenc}
\usepackage{amsmath}
\usepackage{amsfonts}
\usepackage{amssymb}
\usepackage{balance}
\usepackage{graphicx}
\usepackage{epstopdf}
\usepackage{epsfig}
\usepackage{tikz}
\usepackage{soul}
\usetikzlibrary{shapes,arrows}

\usepackage{algorithm}
\usepackage{color}
\usepackage{pifont}% http://ctan.org/pkg/pifont
\usepackage{verbatim,mathrsfs, cite}
\usepackage[acronym,shortcuts]{glossaries}
\usepackage{setspace, url}
\usepackage{bm}

\makeglossaries
\newacronym{5G}{5G}{fifth generation}
\newacronym{AF}{AF}{amplify and forward}
\newacronym{AN}{AN}{artificial noise}
\newacronym{AR}{AR}{auto-regressive}
\newacronym{AUT}{AUT}{authentication}
\newacronym{AWGN}{AWGN}{additive white Gaussian noise}
\newacronym{AKBA}{AKBA}{asymmetric-key based authentication}
\newacronym{BS}{BS}{base station}
\newacronym{BPSK}{BPSK}{binary phase shift keying}
\newacronym{CAS}{CAS}{coding across subchannel}
\newacronym{CDF}{CDF}{cumulative distribution function}
\newacronym{CF}{CF}{compute and forward}
\newacronym{CP}{CP}{cyclic prefix}
\newacronym{CPS}{CPS}{coding per subchannel}
\newacronym{CSIT}{CSIT}{channel-state information at the transmitter}
\newacronym{CSCG}{CSCG}{circularly symmetric complex Gaussian}
\newacronym{DF}{DF}{decode and forward}
\newacronym{DFT}{DFT}{discrete Fourier transform}
\newacronym{FA}{FA}{false alarm}
\newacronym{FD}{FD}{full-duplex}
\newacronym{FBMC}{FMBC}{filterbank modulation}
\newacronym{GSVD}{GSVD}{generalized \ac{SVD}}
\newacronym{GFDM}{GFDM}{generalized frequency division multiplexing}
\newacronym{HD}{HD}{half-duplex}
\newacronym{IDFT}{IDFT}{inverse discrete Fourier transform}
\newacronym{iid}{i.i.d.}{independent and identically distributed}
\newacronym{kkt}{KKT}{Karush-Kuhn-Tucker}
\newacronym{IoT}{IoT}{Internet of things}
\newacronym{LLR}{LLR}{log likelihood ratio}
\newacronym{MAC}{MAC}{multiple access channel}
\newacronym{MAP}{MAP}{maximum a posteriori}
\newacronym{MRC}{MRC}{maximal ratio combining}
\newacronym{MD}{MD}{missed detection}
\newacronym{MI}{MI}{mutual information}
\newacronym{ML}{ML}{maximum likelihood}
\newacronym{MIMO}{MIMO}{multiple-input multiple-output}
\newacronym{MIMOME}{MIMOME}{MIMO multiple-Eve}
\newacronym{MMSE}{MMSE}{minium mean square error}
\newacronym{MSE}{MSE}{mean square error}
\newacronym{MT}{MT}{mobile terminal}
\newacronym{OFDM}{OFDM}{orthogonal frequency division multiplexing}
\newacronym{OFDMA}{OFDMA}{orthogonal frequency division multiple access}
\newacronym{OQAM}{OQAM}{offeset \ac{QAM}}
\newacronym{PDF}{PDF}{probability density function}
\newacronym{PLA}{PLA}{physical layer authentication}
\newacronym{PLS}{PLS}{physical layer security}
\newacronym{PMD}{PMD}{probability mass distribution}
\newacronym{QAM}{QAM}{quadrature amplitude modulation}
\newacronym{RB}{RB}{random binning}
\newacronym{SCF}{SCF}{scaled compute-and-forward}
\newacronym{SC-FDMA}{SC-FDMA}{single-carrier frequency division multiple access}
\newacronym{SKA}{SKA}{secret key agreement}
\newacronym{SKBA}{SKBA}{symmetric-key based authentication}
\newacronym{SKT}{SKT}{secret key throughput}
\newacronym{SWIPT}{SWIPT}{{\em secure} wireless information and power transfer}
\newacronym{SOP}{SOP}{secrecy outage probability}
\newacronym{SVD}{SVD}{singular value decomposition}
\newacronym{SNR}{SNR}{signal to noise ratio}
\newacronym{UFMC}{UFMC}{universal filter multicarrier}
\newacronym{CN}{CN}{complex normal}

\title{Comparison Between Asymmetric and Symmetric Channel-Based Authentication for MIMO Systems}
\author{
	\IEEEauthorblockN{Stefano Tomasin}
	\IEEEauthorblockA{Department of Information Engineering \\ University of Padova, Italy
}}

\begin{document}
\maketitle

\begin{abstract}
Authentication is a key element of security, by which a receiver confirms the sender identity of a message. Typical approaches include either key-based authentication at the application layer or \ac{PLA}, where a message is considered authentic if it appears to have gone through the legitimate channel. In both cases a source of randomness is needed, whereas for \ac{PLA} the random nature of the communication channel is exploited. In this paper we compare the various approaches using in all cases the channel as a source of randomness. We consider a \ac{MIMO} system with a finite number of antennas. Simple \ac{AR} models for its evolution as well as the relation of the legitimate and attacker channel are considered. In this setting the attacker can either predict the key used for key-based authentication or forge the channel estimated at the legitimate receiver for \ac{PLA}. The analysis includes both symmetric and asymmetric key-based authentication. We compare the schemes in terms of false alarm and missed detection probability and we outline best attack strategies.
\end{abstract}


\glsresetall

\section{Introduction} 

By authentication the destination establishes if a received message is coming from the claimed source or not. Typically, this procedure is performed at the application layer by means of cryptographic protocols that are either symmetric or asymmetric, i.e., either a key is shared by the source and destination or a couple of private/public keys is generated by the source that uses the private key to encrypt a message, whose authenticity can be confirmed by any destination having the public key \cite{Stallings}. In both cases, a source of randomness must be available to generate the secret (private) key, and the key must be renewed from time to time to cope with the possibility that the key has been disclosed to the attacker, e.g., due to either protocol vulnerabilities or intensive computational effort by the attacker, possibly supported by quantum computing. 

An alternative authentication approach is implemented (typically in wireless systems) at the physical layer in the so-called \ac{PLA}). It consists in checking if the channel over which the message arrives to the destination remains unaltered over time: in general, the destination will experience different channels to the attacker and the legitimate transmitter, due to their different position and other wireless phenomena. Further details and possible attacks are described in \cite{6204019, 7010914, 7270404}. 

In this paper the unique source of randomness is assumed to be the wireless channel, as estimated by the either or both the transmitter and receiver. Therefore we can compare both \ac{PLA} and cryptography-based authentication techniques. 
We consider a finite number of flat antennas for \ac{MIMO} systems, over which the randomness is obtained to generate the key, therefore we do not apply asymptotic capacity results. The aim is to compare the various authentication schemes in terms of probability of \ac{FA} and \ac{MD}, i.e. the probability that a packet coming from the legitimate source is not authenticated and a packet coming from the attacker is accepted as legitimate, respectively. 

The rest of the paper is organized as follows. Section II introduces the \ac{MIMO} channel model described by a matrix with $N$ entries and its evolution. The three considered authentication schemes, namely \ac{AKBA}, \ac{SKBA} and \ac{PLA} are described in Section III. Performance of the schemes in terms of \ac{FA} and \ac{MD} probabilities is assessed in Section IV. Finally, conclusions are outlined in Section V.

\section{System Model}

We consider the usual three users model in security, where Alice aims at authenticating messages coming from Bob, and Eve aims at sending messages to Alice impersonating Bob. In particular, after an initialization stage, upon reception of a packet Alice aims at taking a decision between the two hypothesis $\mathcal H_0$: the packet is coming from Bob or $\mathcal H_1$: the packet is not coming from Bob.

We assume that the only sources of randomness are the time-varying channels among devices, which are estimated and used to provide the desired authentication process. In particular, devices are equipped with $M$ antennas each, implementing a \ac{MIMO} system, so that channels are described by $M \times M$ matrices. We assume that entries of channel matrices are \ac{iid} \ac{CN}, and correlated over time and across devices. 

Channel matrices are converted into vectors of size $N =M^2$, so that $\bm{h}(t) = [h_1(t), \ldots, h_N(t)]$ is the Alice-Bob channel (assumed to be reciprocal) at slot $t$. We assume an \ac{AR} evolution of the channel, i.e., for $n=1, \ldots, N$
\begin{equation}
h_n(t) = \alpha h_n(t-1) + \sqrt{1-\alpha^2} z_n(t)\,,
\label{AR}
\end{equation}
where $\alpha$ is the correlation factor and $z_n(t)$ are \ac{iid} \ac{CN} variables. 

A similar channel model is available between Eve and both Alice and Bob. In particular the reciprocal Alice-Eve channel is 
\begin{equation}
g_{1, n}(t) = \beta_1 h_n(t) + \sqrt{1 - \beta_1^2} z_{1, n}(t)\,,
\label{eqg1}
\end{equation}
and the Bob-Eve channel is
\begin{equation}
g_{2, n}(t) = \beta_2 h_n(t) + \sqrt{1 - \beta_2^2} z_{2, n}(t)\,,
\end{equation}
where $\beta_\cdot$ are correlation factors and $z_{\cdot, n}(t)$ are \ac{CN} variables.

In order to exploit the channel as a random source the users must estimate it, and this is obtained by letting one of the two legitimate users to transmit pilot symbols. In particular, the channel estimate obtained by Alice when Bob is transmitting pilots is an \ac{AWGN}-corrupted version of $h_n(t)$, i.e.,
\begin{equation}
\hat{h}_{{\rm A}, n}(t) = h_n(t) + \sigma_{\rm A} w_{\rm A, n}(t)\,,
\end{equation}
where $w_{\rm A, n}(t)$ are \ac{iid} \ac{CN} variable. Similarly for a pilot transmission by Alice, Bob estimates 
\begin{equation}
\hat{h}_{\rm B, n}(t) = h_n(t) + \sigma_{\rm B} w_{\rm B, n}(t)\,,
\end{equation}
where $w_{\rm B, n}(t)$ are \ac{iid} \ac{CN} variable. Lastly, when either Alice ($i=1$) or Bob ($i=2$) are transmitting, Eve obtains estimates
\begin{equation}
\hat{g}_{i,n}(t) = g_{i,n}(t) + \sigma_{\rm E} w_{\rm E, n}(t)\,,
\end{equation}
where $w_{\rm E, n}(t)$ are \ac{iid} \ac{CN} variable.

\paragraph*{Attacker Model} Here we consider that Eve will only try to transmit packets impersonating Alice, while not performing attacks aiming at disrupting the authentication process (such as jamming and pilot contamination attacks) that will be considered in future studies.  We assume that Eve is able to transmit modifying the transmitted signal so that Alice will estimate any desired channel.

%
%
%
%
% Then we define
%\begin{equation}
%g_{1,n}(t) = \frac{q_{1, n}(t)}{\beta_1 } = \frac{h_n(t)}{\sigma_1} + z_{1,n}(t)\,,
%\end{equation}
%\begin{equation}
%g_{2,n}(t) = \frac{q_{2, n}(t)}{\beta_2 } = \frac{h_n(t)}{\sigma_2} + z_{2,n}(t)\,,
%\end{equation}
%where $z_{i,n}(t)$ are zero-mean unit-variance complex Gaussian variables while 
%\begin{equation}
%\sigma_i^2 = \frac{2 - \beta_1^2}{\beta_i^2}\,.
%\end{equation}
%
%
%
%The {attacker model} provides that a) for {\em key-based schemes} he predicts the secret keys using estimates of $g_{1,n}(t)$ and $g_{2,n}(t)$ and uses it to sends messages or b) for {\em \ac{PLA}} he estimates the channel $h_n(t)$ by using $g_{1,n}(t)$ and $g_{2,n}(t)$ and then he {\em forges} the channel that is going to be estimated by the legitimate receiver (see \cite{6204019} for details). 
%
%Estimates of $h_n(t)$

\section{Authentication Schemes}

Three authentication schemes are considered: \ac{PLA}, \ac{AKBA} and \ac{SKBA}.

\subsection{Physical Layer Authentication} 

\ac{PLA} exploits the coherence time of the channel in order to provide authentication. In synthesis, it provides two steps: a) the initialization step by which the channel to the user to be authenticated is measured, being sure that the legitimate node is transmitting; b) the authentication step, in which upon reception of a packet the channel is estimated again and compared with that estimated in the initialization step. 

Let us now consider the detailed implementation of \ac{PLA} in the considered scenario. We first observe that we can rewrite $\hat{h}_{{\rm A}, n}(t)$ as 
\begin{equation}
\hat{h}_{{\rm A}, n}(t) =  \alpha^{t-1} h_n(1) + \gamma_{\rm A}(t) \epsilon_n(t)\,,
\end{equation}
with $\epsilon_n(t)$ \ac{iid} \ac{CN} variables and $\gamma_{\rm A}^2(t) = \sigma_{\rm A}^2 + (1-\alpha^2)\alpha^{t-1}$.

The physical layer authentication as described in \cite{6204019} includes the following phases:
\begin{enumerate}
\item In the first slot Bob transmits pilots to Alice, who estimates the channel $\hat{h}_{{\rm A}, n}(1)$. This first transmission is assumed to be authenticated with some other means other than \ac{PLA}.
\item At slot $t > 1$, when Alice receives a packet that contains pilot symbols, she estimates the channel $\hat{h}_{{\rm A}, n}(t)$ then she computes the distance between it and the channel estimated at the first slot as 
\begin{equation}
\psi(t) = \frac{1}{N\gamma^2_{\rm A}(t)} \sum_{n=1}^N |\hat{h}_{{\rm A}, n}(t) - \alpha^{t-1}\hat{h}_{{\rm A}, n}(1)|^2\,,
\end{equation}
and decides that the packet is authentic if 
\begin{equation}
\psi(t) < \theta(t)\,,
\end{equation}
otherwise she discards it as non-authentic.
\item Eve on her side estimates the channel from Bob in the first slot $\hat{g}_{2, n}(1)$ and in forthcoming slots she estimates the channel from Alice and Bob whenever they are transmitting. Then she obtain the \ac{ML} estimate of the Alice-Bob channel at the first slot $\tilde{h}_{{\rm E},n}(t)$, which is used to impersonate Bob by forging channel $\alpha^{t-1}\tilde{h}_{{\rm E},n}(t)$ at slot $t$.
\end{enumerate}

We now suppose that at odd slots (starting from $t=1$) Bob is transmitting, and at even slots Alice is transmitting. Then at slot $t = 2\nu+1$, with $\nu$ natural number, Eve performs the attack. From (\ref{AR}) and (\ref{eqg1}) we can write $\hat{g}_{i,n}(t)$ as a function of $h_n(1)$ as 
\begin{equation}
\begin{split}
\hat{g}_{i,n}(t) = & \beta_i \alpha^{t-1} h_n(1) + \beta_i\sqrt{1-\alpha^2}  \sum_{k=2}^t \alpha^{t-k} z_n(k) + \\
& + \sqrt{1-\beta_i^2} z_{i,n}(t) + \sigma_{\rm E}w_{{\rm E},n}(t)\\
= & \beta_i \alpha^{t-1} h_n(1) + u_{i,n}(t)\,,
\end{split}
\label{nuovagh}
\end{equation}
with $u_{i,n}(\tau)$, $\tau=1, \ldots, t$ \ac{CN} variables. Now, let $\bm{w}(2\nu)$ be a $2\nu$-size column vector with entries $[\bm{w}]_{2j+i} = \beta_{2-i} \alpha^{2j+i-1}$, and $\hat{\bm{g}}_n(2\nu)$ be the column vector collecting channel estimates at Eve up to slot $2\nu$ for channel entry $n$, then
\begin{equation}
\hat{\bm{g}}_n(2\nu) = \bm{w} h_n(1) + \bm{u}\,,
\end{equation}
with Gaussian zero-mean random vector $\bm{u}$ having correlation matrix $\bm{K} = \mathbb E[\bm{u}^H \bm{u}]$ with entries 
\begin{equation}
\begin{split}
[\bm{K}&]_{t_1,t_2} =   (\sigma_{\rm E}^2 + 1 - \beta_{1+t_1|_2)}^2)\delta(t_1-t_2)  \\
 & +\beta_{1+t_1|_2}\beta_{1+t_2|_2}(1-\alpha^2) \times \\
 & \times \mathbb E\left[   \sum_{k_1=2}^{t_1} \alpha^{t_1-k_1} z_n(k_1)  \sum_{k_2=2}^{t_2} \alpha^{t_2-k_2} z_n(k_2)\right] \\
 =&   (\sigma_{\rm E}^2 + 1 - \beta_{1+t_1|_2}^2)\delta(t_1-t_2) \\
 & + \beta_{1+t_1|_2}\beta_{1+t_2|_2}(1-\alpha^2)  \sum_{k=2}^{\min\{t_1, t_2\}}  \alpha^{2(t-k)} \\
= &  (\sigma_{\rm E}^2 + 1 - \beta_{1+t_1|_2}^2)\delta(t_1-t_2) \\
 & + \beta_{1+t_1|_2}\beta_{1+t_2|_2}(1 - \alpha^{2(\min\{t_1, t_2\}-1)})\,,
\end{split}
\end{equation}
and $t|_2$ denotes the reminder of a division of $t$ by 2. The \ac{ML} estimate of $h_n(1)$ is obtained as
\begin{equation}
\tilde{h}_{{\rm E},n}(t)  = (\bm{w}^H\bm{K}^{-1} \bm{w})^{-1}\bm{w}^H\bm{K}^{-1}\hat{\bm{g}}_n(2\nu)\,.
\label{MLeve}
\end{equation}

When multiple attacks are possible, an exploration of the channel space $\bm{h}(1)$ can be performed, where the first attempted point is the \ac{ML} estimate (\ref{MLeve}) and then the other points correspond to the next most probable channels, given the observations. This approach has been explored in \cite{6204019}.

\subsection{Asymmetric-key Based Authentication} 

With \ac{AKBA} we exploit the channel as a random number generator and the random number is then used to generate a couple of private and public keys that will be used to sign and check the packets to be authenticated. In particular, the randomness is used by Bob to generate a private/public key couple. The public key is broadcast to all users and when Bob transmits packets he will encrypt a time-varying signature with the private key, and Alice will be able to decrypt it using the public key, thus ensuring that the packet is authentic.

Let us now provide the details of key generation for \ac{AKBA} in the considered scenario. The \ac{AKBA} works as follows:
\begin{enumerate}
\item At slot 1 Alice transmits pilots and Bob estimates the channel $\hat{h}_{{\rm B}, n}(1)$. 
\item Bob quantizes the real and imaginary parts of the channel estimate with a quantizer of $K_{\rm Q}$ quantization levels and saturation value $v_{\rm sat}$, obtaining a bit sequence $\bm{b}$.
\item Bob applies a known (to all users) hashing and compression function to extract a shorter key from $\bm{b}$ and then generate a private and public key couple ($\bm{S}_{\rm A}$, $\bm{P}_{\rm A}$).
\item At slot 2 Bob broadcasts the public key $\bm{P}_{\rm A}$. In his transmissions Bob encodes a (time-varying) signature with the private key so that Alice is able to authenticate messages coming from Bob.
\item Eve estimates the channel in the first slot $\hat{g}_{1,n}(1)$ and the channel in the second slot $\hat{g}_{2,  n}(2)$.
\item From the two estimates she will obtain a bit sequence $\hat{\bm{b}}$ as detailed in the following.
\item Eve uses $\hat{\bm{b}}$ as the key to generate a private and public key couple ($\bm{S}_{\rm E}$, $\bm{P}_{\rm E}$) and performs attacks signing the time-varying signature with the private key $\bm{S}_{\rm E}$.
\end{enumerate}

\paragraph*{Optimal Attack} Let us now derive the optimal choice of $\hat{\bm{b}}$ by Eve at slot $t = 2\nu+1$ after she has collected $\nu$ observations of the channels to Alice and Bob. We now have that at odd slots (starting from $t=1$) Bob is transmitting, and at even slots Alice is transmitting. Now, let $\bm{w}'$ be a $2\nu$-size column vector with entries $[\bm{w}']_{2j+i} = \beta_i \alpha^{2j+i-1}$, and let $\hat{\bm{g}}'_n(2\nu)$ be the column vector collecting channel estimates at Eve up to slot $2\nu$ for channel entry $n$, then
\begin{equation}
\hat{\bm{g}}'_n(2\nu) = \bm{w}' h_n(1) + \bm{u}'\,,
\end{equation}
where Gaussian zero-mean random vector $\bm{u}'$ has correlation matrix $\bm{K}' = \mathbb E[\bm{u}^{'H} \bm{u}']$ with entries (obtained analogously as in the previous section)
\begin{equation}
\begin{split}
[\bm{K}'&]_{t_1,t_2} =  (\sigma_{\rm E}^2 + 1 - \beta_{2-t_1|_2}^2)\delta(t_1-t_2) \\
& + \beta_{2-t_1|_2}\beta_{2-t_2|_2}(1 - \alpha^{2(\min\{t_1, t_2\}-1)})\,.
\end{split}
\end{equation}

Indicating with $(\tau_k, \tau_{k+1}]$ the quantization interval for the quantization level $k = 1, \ldots, K_{\rm Q}$, and $\tau_1 = - \tau_{K_{\rm Q}+1} = \infty$, the index of the most probable quantized value (for the real part) is 
\begin{equation}
\begin{split}
k^*  & = {\rm argmax}_k \mathbb P[\Re\{h_n(1)\} \in (\tau_k, \tau_{k+1}]|\hat{\bm{g}}'_n(2\nu)]  \\
= &   {\rm argmax}_k \int_{\tau_k}^{\tau_{k+1}} p_{\Re\{h_n(1)\} |\hat{\bm{g}}'_n(2\nu)}(h |\hat{\bm{g}}'_n(2\nu)) {\rm d} h \\
 =&   {\rm argmax}_k \int_{\tau_k}^{\tau_{k+1}} p_{\hat{\bm{g}}'_n(2\nu)|\Re\{h_n(1)\}}(\hat{\bm{g}}'_n(2\nu)|h )  {\rm d} h  \\
 = & {\rm argmax}_k \int_{\tau_k}^{\tau_{k+1}} e^{-\frac{1}{2}(\hat{\bm{g}}'_n(2\nu) - \bm{w}'h)^H\bm{K}^{-1}(\hat{\bm{g}}'_n(2\nu) - \bm{w}'h)} {\rm d} h\,.
\end{split}
\end{equation}
Now, defining 
\begin{gather}
a = \bm{w}^{'H}\bm{w}'\,, \\
b = -\hat{\bm{g}}^{'H}_n(2\nu)\bm{K}'\bm{w}' - \bm{w}^{'H}\bm{K}'\hat{\bm{g}}'_n(2\nu)\,, 
\end{gather}
we have
\begin{equation}
\begin{split}
k^*  & =  {\rm argmax}_k \int_{\tau_k}^{\tau_{k+1}} e^{-(ah^2+bh)} {\rm d} h \\
& = {\rm argmax}_k  \left[{\rm erf}\left(\frac{2a \tau_{k+1} +b}{2\sqrt{a}} \right) - {\rm erf}\left(\frac{2a \tau_k +b}{2\sqrt{a}} \right)\right]\,,
\end{split}
\label{kstar}
\end{equation}
with ${\rm erf}$ the error function of the normal distribution.

Also in this case if multiple attacks are available, various quantized points can be explored, jointly among all $2N$ quantized value. Therefore, after having considered the key obtained using quantized values (\ref{kstar}), we must find the next quantized values in all $N$ observations having the highest probability, thus changing only one quantized value, and so on.

\subsection{Symmetric-key Based Authentication}

With \ac{SKBA} the two legitimate devices must agree on a secret key by which authentication is performed. In particular, once Alice and Bob has agreed on a (secret) key, when Bob is transmitting a packet it will encode a time-varying signature with the secret key so that Alice can check that the packet is coming from Bob.

For the establishment of the secret key we resort to the source-based secret key agreement method \cite{Bloch} where the two users extract the quantized randomness from the channel and then go through the steps of advantage distillation, information reconciliation and privacy amplification. In particular, in the step of information reconciliation error correcting codes are used to ensure that the keys between the two users coincide. We consider the following scheme, in which all users have agreed on a codebook $\mathcal C$ of codewords of length $N$:
\begin{enumerate}
\item At slot 1 Alice transmits a set of pilot symbols and Bob estimates the channel $h_{{\rm B},n}(1)$. Eve estimates the channel $\hat{g}_{1,n}(1)$.
\item At slot 2 Bob transmits a set of pilot symbols and Alice estimates the channel $h_{{\rm A},n}(2)$. Alice finds in $\mathcal C$ the codeword closest to $\bm{h}_{\rm A}(2)$, vector collecting the channels for all $2N$ observations , i.e.
\begin{equation}
\bm{c}_{\rm A}^* = {\rm argmin}_{\bm{c} \in \mathcal C} ||\hat{\bm{h}}_{\rm A}(2) - \bm{c}||^2\,.
\end{equation} 
Eve estimates the channel $\hat{g}_{2, n}(2)$.
\item Alice sends the error of her estimated channel with respect to the selected codeword, i.e.,
\begin{equation}
\bm{\epsilon} = \hat{\bm{h}}_{\rm A}(2) - \bm{c}^*_{\rm A}\,.
\end{equation}
\item Bob computes $\tilde{\bm{h}}_{\rm B} = \hat{\bm{h}}_{\rm B}(1) - \bm{\epsilon}$ and decodes it in the codebook $\mathcal C$, i.e., computes
\begin{equation}
\bm{c}_{\rm B}^* = {\rm argmin}_{\bm{c} \in \mathcal C} ||\tilde{\bm{h}}_{\rm B} - \bm{c}||^2\,.
\end{equation} 
\item Eve obtains from the two estimates at point 1) and 2) the \ac{ML} estimate of the Alice-Bob channel $\tilde{\bm{h}}_{\rm E}(1)$, adds $\bm{\epsilon}$ and decodes the resulting vector, i.e., she computes 
\begin{equation}
\bm{c}_{\rm E}^* = {\rm argmin}_{\bm{c} \in \mathcal C} ||\tilde{\bm{h}}_{\rm E}(1) - \bm{\epsilon} - \bm{c}||^2\,.
\end{equation} 
\item The three users will apply a hash function to the index of the decoded sequences to extract $R_n$ bits that will be used as key for authentication.
\end{enumerate}

In this case multiple attacks can be performed by considering the next most probable decoded codewords, given the observations.

\section{Performance Analysis}

We consider now the performance of each authentication scheme separately, in terms of \ac{FA} and \ac{MD} probabilities.

\subsection{Physical Layer Authentication} 

For the \ac{PLA} scheme \ac{FA} and \ac{MD} probabilities have been derived in \cite{6204019} without considering the channel evolution of  (\ref{AR}).  

Conditioned on $\mathcal H_0$ and for any value of $\hat{h}_{{\rm A},n}(1)$, $\Psi(t)$, $t > 1$ is a central chi-square distributed random variable with $2n$ degrees of freedom, yielding the \ac{FA} probability   
\begin{equation}
P_{\rm PLA, FA} = {\rm P}[\Psi(t) > \theta(t)| \mathcal{H}_{0}]  = 1 - F_{2n,0}(\theta(t))\,,
\label{pFA}
\end{equation}
where 
\begin{equation}
\begin{split}
F_{2n,\mu}(x)=& 1-{\rm Q}_n(\sqrt{\mu}, \sqrt{x})\\
 = &e^{-\frac{\mu}{2}} \sum_{k=0}^{\infty} \left(-\frac{\mu}{2} \right)^k \frac{\gamma\left(n + k, \frac{x}{2}\right)}{k!}
 \end{split}
\end{equation}
is the \ac{CDF} of a chi-square random variable with $2n$ degrees of freedom and noncentrality parameter $\mu$, ${\rm Q}_n(\sqrt{\mu}, \sqrt{x})$ is the Marcum Q-function, and $\gamma(a;b)$ is the normalized lower incomplete gamma function.

Conditioned on $\mathcal H_1$, the realization of $\hat{h}_{{\rm  A}_n}(1)$ and the forged channel $\tilde{h}_{{\rm E},n}(1)$, $\Psi(t)$ is a noncentral chi-square distributed random variable with $2n$ degrees of freedom and noncentrality parameter 
\begin{equation}
\beta(t) = \frac1{N\gamma_{\rm A}^2(t)} \sum_{n=1}^N |\tilde{h}_{{\rm E},n}(t) - \alpha^{t-1}\hat{h}_{{\rm  A},n}(1) |^2 
\label{beta}
\end{equation}
which provides the \ac{MD} probability
\begin{equation}
\begin{split}
P_{\rm PLA, MD}(\bm{h}(1), \tilde{\bm{h}}(t))  & = \mathbb P[\Psi(t) \leq \theta(t)| \mathcal{H}_{1}, \bm{h}(1), \tilde{\bm{h}}(t)]  \\
& = F_{2n,\beta(t)}(\theta(t))\,.
\end{split}
\label{pMD_hg}
\end{equation}

\subsection{Asymmetric-key Based Authentication} 

For the \ac{AKBA} scheme, Bob generates a secret private key from the channel and then he obtains a public key. Assuming that the broadcast of the public key is error-free and authenticated, no \ac{FA} will ever occur, thus 
\begin{equation}
P_{\rm AKBA, FA} = 0\,.
\end{equation}
The \ac{MD} will instead be related to the ability of Eve to predict the private key.  In particular we have
\begin{equation}
P_{\rm AKBA, MD} = \mathbb P[\bm{b} = \hat{\bm{b}}] = \mathbb P[\bm{q} = \hat{\bm{q}}]\,,
\end{equation}
where $\bm{q}$ and $\hat{\bm{q}}$ are vectors of quantized values by Alice and Eve, respectively.

\subsection{Symmetric-key Based Authentication} 

For the \ac{SKBA} scheme the \ac{FA} probability is the probability that $\bm{c}^*_{\rm A}$ and $\bm{c}^*_{\rm B}$ do not coincide. The \ac{MD} probability is the probability that $\bm{c}^*_{\rm E}$ and $\bm{c}^*_{\rm A}$  coincide, so that Eve is able to break the authentication system. 

We first observe that all users have $\bm{\epsilon}$ available and the resulting estimates after its removal are
 \begin{gather}
\tilde{\bm{h}}_{\rm A} = \bm{c}^*_{\rm A}\,, \quad \tilde{\bm{h}}_{\rm B} = \bm{c}^*_{\rm A} + \sigma_{\rm B}\bm{w}_{\rm B} + \sigma_{\rm A}\bm{w}_{\rm A }\,, \\
\tilde{\bm{h}}_{\rm E} = \bm{c}^*_{\rm A} + \sigma_{\rm E}\bm{w}_{\rm E} + \sigma_{\rm A}\bm{w}_{\rm A}\,.
\end{gather}
Hence the \ac{FA} probability $P_{\rm SKBA, FA}$ is the probability that Bob does not decode $\bm{c}^*_{\rm A}$ and the \ac{MD} probability  $P_{\rm SKBA, MD}$ is the probability that Eve decodes $\bm{c}^*_{\rm A}$. 

Considering the hypothesis testing problem of decoding at Bob, we define the log-likelihood function
\begin{equation}
\Lambda_{\rm B}(\bm{c}^*_{\rm A},\tilde{\bm{h}}_{\rm B}) = \frac{1}{N} \ln \frac{p_{\tilde{\bm{h}}_{\rm B}|\tilde{\bm{h}}_{\rm A}}(\tilde{\bm{h}}_{\rm B}|\bm{c}^*_{\rm A})}{p_{\tilde{\bm{h}}_{\rm B}}(\tilde{\bm{h}}_{\rm B})}\,, 
\end{equation}
and we have that $P_{\rm SKBA, FA}$ is the {\em missed detection} probability of the detection process, i.e., 
\begin{equation}
\begin{split}
P&_{\rm SKBA, FA} =  \pi_{\rm B, MD}(\bm{c}^*_{\rm A}, \lambda)  \\
= & \mathbb P[\Lambda_{\rm B}(\bm{c}^*_{\rm A},\tilde{\bm{h}}_{\rm B}) \leq \lambda]\,, \quad \tilde{\bm{h}}_{\rm B} \sim p_{\tilde{\bm{h}}_{\rm B}|\tilde{\bm{h}}_{\rm A}}, \tilde{\bm{h}}_{\rm A} = \bm{c}_{\rm A}^*\,.
\end{split}
\end{equation}
On the other hand, the \ac{FA} probability of the detection process sets an upper bound on the rate of the secret key $R$, therefore the threshold $\lambda$, in particular
\begin{equation}
R \leq -\frac{1}{N}\log_2 \pi_{\rm B, FA}(\lambda)
\end{equation}
where
\begin{equation}
\pi_{\rm B, FA}(\lambda) = \mathbb P[\Lambda_{\rm B}(\bm{c}^*_{\rm A},\tilde{\bm{h}}_{\rm B}) > \lambda] \quad  \tilde{\bm{h}}_{\rm B} \sim p_{\tilde{\bm{h}}_{\rm B}}\,.
\end{equation}
Following similar derivations to those for the \ac{FA}, it turns out that the \ac{MD} probability of the authentication process is the complementary of the \ac{MD} probability of the detection process at Eve, i.e., defining 
\begin{equation}
\Lambda_{\rm E}(\bm{c}^*_{\rm A},\tilde{\bm{h}}_{\rm E}) = \frac{1}{N} \ln \frac{p_{\tilde{\bm{h}}_{\rm E}|\tilde{\bm{h}}_{\rm A}}(\tilde{\bm{h}}_{\rm E}|\bm{c}^*_{\rm A})}{p_{\tilde{\bm{h}}_{\rm E}}(\tilde{\bm{h}}_{\rm E})}\,, 
\end{equation}
we have
\begin{equation}
\begin{split}
P&_{\rm SKBA, MD} =  1 - \pi_{\rm B, MD}(\bm{c}^*_{\rm A}, \lambda) \\
= & 1 - \mathbb P[\Lambda_{\rm E}(\bm{c}^*_{\rm A},\tilde{\bm{h}}_{\rm E}) > \lambda]\,, \quad \tilde{\bm{h}}_{\rm E} \sim p_{\tilde{\bm{h}}_{\rm E}|\tilde{\bm{h}}_{\rm A}}, \tilde{\bm{h}}_{\rm A} = \bm{c}_{\rm A}^*\,.
\end{split}
\end{equation}
Efficient methods to compute the \ac{FA} and \ac{MD} probabilities of the detection process have been derived in \cite{Erseghe1}.

\balance
\section{Conclusions}

In this paper we have considered a \ac{MIMO} time-varying channel model among two legitimate users and an eavesdropper. We have compared three authentication methods, all based on the randomness of the \ac{MIMO} channels: \acl{PLA}, \acl{AKBA} and \acl{SKBA}. We have described how to exploit the channel in this scenario, which is the best attack by the eavesdropper and we have obtained the performance in terms of \ac{FA} and \ac{MD} probabilities for the authentication process. 

\bibliographystyle{IEEEtran}

\bibliography{../references}

\begin{thebibliography}{1}
\providecommand{\url}[1]{#1}
\csname url@samestyle\endcsname
\providecommand{\newblock}{\relax}
\providecommand{\bibinfo}[2]{#2}
\providecommand{\BIBentrySTDinterwordspacing}{\spaceskip=0pt\relax}
\providecommand{\BIBentryALTinterwordstretchfactor}{4}
\providecommand{\BIBentryALTinterwordspacing}{\spaceskip=\fontdimen2\font plus
\BIBentryALTinterwordstretchfactor\fontdimen3\font minus
  \fontdimen4\font\relax}
\providecommand{\BIBforeignlanguage}[2]{{%
\expandafter\ifx\csname l@#1\endcsname\relax
\typeout{** WARNING: IEEEtran.bst: No hyphenation pattern has been}%
\typeout{** loaded for the language `#1'. Using the pattern for}%
\typeout{** the default language instead.}%
\else
\language=\csname l@#1\endcsname
\fi
#2}}
\providecommand{\BIBdecl}{\relax}
\BIBdecl

\bibitem{Stallings}
W.~Stallings, \emph{Cryptography and Network Security: Principles and
  Practice}, 5th~ed.\hskip 1em plus 0.5em minus 0.4em\relax Upper Saddle River,
  NJ, USA: Prentice Hall Press, 2010.

\bibitem{6204019}
P.~Baracca, N.~Laurenti, and S.~Tomasin, ``Physical layer authentication over
  mimo fading wiretap channels,'' \emph{IEEE Transactions on Wireless
  Communications}, vol.~11, no.~7, pp. 2564--2573, July 2012.

\bibitem{7010914}
A.~Ferrante, N.~Laurenti, C.~Masiero, M.~Pavon, and S.~Tomasin, ``On the error
  region for channel estimation-based physical layer authentication over
  {Rayleigh} fading,'' \emph{IEEE Transactions on Information Forensics and
  Security}, vol.~10, no.~5, pp. 941--952, May 2015.

\bibitem{7270404}
E.~Jorswieck, S.~Tomasin, and A.~Sezgin, ``Broadcasting into the uncertainty:
  Authentication and confidentiality by physical-layer processing,''
  \emph{Proceedings of the IEEE}, vol. 103, no.~10, pp. 1702--1724, Oct 2015.

\bibitem{Bloch}
M.~Bloch and J.~Barros, \emph{Physical-Layer Security. From Information Theory
  to Security Engineering}.\hskip 1em plus 0.5em minus 0.4em\relax Cambridge:
  Cambridge University Press, 2011.

\bibitem{Erseghe1}
\BIBentryALTinterwordspacing
T.~Erseghe, ``Coding in the finite-blocklength regime: Bounds based on
  {Laplace} integrals and their asymptotic approximations,'' \emph{CoRR}, vol.
  abs/1511.04629, 2015. [Online]. Available:
  \url{http://arxiv.org/abs/1511.04629}
\BIBentrySTDinterwordspacing

\end{thebibliography}

\end{document}